\documentclass[]{aa}
\usepackage{multirow}
\usepackage{gensymb}
\usepackage{subfigure}
\usepackage{graphicx}
\usepackage{natbib}
\usepackage{appendix}
\usepackage[english]{babel}
\usepackage{txfonts}
\usepackage{footnote}

\makeatletter
\newcommand*{\myfnsymbolsingle}[1]{%
  \ensuremath{%
    \ifcase#1
    \or 
      *%
    \or 
      \dagger
    \or 
      \ddagger
    \or 
      \mathsection
    \or 
      \mathparagraph
    \else 
      \@ctrerr  
    \fi
  }%
}   
\makeatother

\usepackage{alphalph}
\newalphalph{\myfnsymbolmult}[mult]{\myfnsymbolsingle}{}

\begin{document}

\newcommand{\Msun}{M$_{{\odot}}$\:}
\newcommand{\pow[1]}{10$^{#1}$}
\newcommand{\Rin}{R$_{\mathrm{in}}$\:}
\newcommand{\Rout}{R$_{\mathrm{out}}$\:}
\newcommand{\kms}{km~s$^{-1}$}
\newcommand{\Lsun}{L$_{{\odot}}$}

\titlerunning{High spatial resolution imaging of SO and H$_2$CO in AB Auriga: The
first SO image in a transitional disk}
\authorrunning{Pacheco-V\'azquez et al. 2016}

\title{High spatial resolution imaging of SO and H$_2$CO in AB Auriga:\\The
first SO image in a transitional disk\footnotemark}
\subtitle{}

   \author{ S. Pacheco-V\'azquez\inst{1},
	    A. Fuente\inst{1},
	    C. Baruteau\inst{2},
	    O. Bern\'e\inst{2,3},
	    M. Ag\'undez \inst{4},
	    R. Neri\inst{5},
            J. R. Goicoechea\inst{4},
            J. Cernicharo\inst{4},
            \and
            R. Bachiller\inst{1}
          }

   \institute{
     Observatorio Astron\'omico Nacional (OAN), Apdo 112, E-28803 Alcal\'a de Henares, Madrid, Spain\\
          \email{s.pacheco@oan.es, a.fuente@oan.es}\and
     CNRS, IRAP, 9 Av. colonel Roche, BP 44346, F-31028 Toulouse cedex 4, France\and
     Université de Toulouse, UPS-OMP, IRAP, Toulouse, France\and
     Instituto de Ciencia de Materiales de Madrid, ICMM-CSIC, C/ Sor Juana Inés de la Cruz 3, E-28049 
     Cantoblanco, Spain\\
          \email{marcelino.agundez@icmm.csic.es}\and
     Institut de Radioastronomie Millim\'etrique, 300 Rue de la Piscine, F-38406 Saint Martin d'Hères, 
     France\\
             }

   \date{}

\abstract
  {Transitional disks are structures of dust and gas around young stars 
with large inner cavities in which planet formation may occur. Lopsided 
dust distributions are observed in the dust 
continuum emission at millimeter wavelengths. These asymmetrical 
structures can be explained as being the result of an enhanced gas density vortex
where the dust is trapped, potentially promoting the rapid growth to 
the planetesimal scale.}
   {AB Aur hosts a transitional disk with a clear horseshoe morphology
which strongly suggests the presence of a dust trap. Our goal is to 
investigate its formation and the possible effects on the gas chemistry.}
   {We used the NOEMA (NOrthern Extended Millimeter Array) interferometer to image the 1mm 
continuum dust emission and the $^{13}$CO 
J$=$2 $\rightarrow$1, C$^{18}$O J$=$2 $\rightarrow$1, 
SO J$=$5$_6$ $\rightarrow$4$_5$, and H$_2$CO J$=$3$_{03}$ $\rightarrow$2$_{02}$ 
rotational lines.}
   {Line integrated intensity ratio images are built to investigate the chemical
changes within the disk. 
The I(H$_2$CO J$=$3$_{03}$ $\rightarrow$2$_{02}$)/I(C$^{18}$O J$=$2$\rightarrow$1) 
ratio is fairly constant along the disk with values of $\sim$0.15$\pm$0.05. On 
the contrary, 
the I(SO J$=$5$_6$ $\rightarrow$4$_5$)/I(C$^{18}$O J$=$2 $\rightarrow$1) and 
I(SO J$=$5$_6$ $\rightarrow$4$_5$)/I(H$_2$CO J$=$3$_{03}$ $\rightarrow$2$_{02}$)
ratios present a clear northeast-southwest gradient 
(a factor of 3$-$6) with the minimum towards the dust trap. This gradient cannot 
be explained
by a local change in the excitation conditions but by a decrease in the SO
abundance. Gas densities up to $\sim$10$^9$ cm$^{-3}$ are expected in the disk midplane
and two-three times larger in the high pressure vortex. We have used a 
single point (n,T) chemical model to investigate the lifetime of gaseous CO, H$_2$CO, and SO in
the dust trap. Our model shows that for densities $>$10$^7$~cm$^{-3}$, 
the SO molecules are depleted 
(directly frozen, or converted into SO$_2$ and then frozen out)
in less than 0.1~Myr. The lower SO abundance towards the dust trap could indicate that a larger fraction of the gas is in a high density environment.}
   {Gas dynamics, grain growth and gas chemistry are coupled in the planet formation process. We detect a chemical
signature of the presence of a dust trap in a transitional disk. Because of the strong dependence of SO abundance on the gas density,
the sulfur chemistry can be used as a chemical diagnostic to detect the birthsites of future planets. However, the large
uncertainties inherent to chemical models and the limited knowledge of the disk's physical structure and initial conditions are important drawbacks.}
   
   \keywords{stars: formation -- stars: individual: AB Aur -- stars: pre-main
sequence -- stars: variables: T Tauri, Herbig Ae/Be -- circumstellar matter --
protoplanetary disks}
    \maketitle

\footnotetext{$^\star$ Based on observations carried out under project number S14AO with the IRAM NOEMA Interferometer. IRAM is supported by INSU/CNRS (France), MPG (Germany) and IGN (Spain).}

\section{Introduction}

\begin{table*}[]
\centering
\caption{NOEMA interferometer map parameters.}
\begin{tabular}{p{2.1cm}p{1.5cm}ccccp{1.4cm}p{0.5cm}}\hline\hline
Species       &Freq.     &Beam              &PA      &E$_{\rm u}$& A$_{\rm ul}$  &Noise   &$\Delta$v\\
              &GHz     &arcsec          &$^{\circ}$&K&s$^{-1}$ &Jy~beam$^{-1}$&km~s$^{-1}$\\\hline
$^{13}$CO 2-1&220.398684&1.64$\times$1.53  &1       &15.9&6.08e-07 &2.3e-02       &0.2\\
C$^{18}$O 2-1&219.560357 &1.64$\times$1.53 &1       &15.8&6.01e-07 &2.3e-02  &0.2\\
H$_2$CO 3$_{03}$-2$_{02}$ &218.222192  &1.65$\times$1.55 &3&21.0&2.82e-04    &3.1e-02  &0.1 \\
SO 5$_6$-4$_5$&219.949442&1.64$\times$1.53  &1      &35.0&1.36e-04           &3.2e-02  &0.1\\
\hline         
\end{tabular}
\label{tab:PdBI}
\end{table*}

Transitional disks are objects around young stars with large cavities cleared of small dust grains in the inner disk regions.
The formation of planetesimals requires that primordial dust grains grow from micron- to km-sized bodies. 
Lopsided dust distributions have been commonly identified in transitional disks using 
the dust continuum emission at millimeter wavelengths \citep{Espaillat2014}. 
It is possible that the size distribution of the dust grains is not uniform throughout the disk: micron- and millimeter-sized grains may have different 
distributions \citep{Marel2013,Marel2015,Pinilla2015c}. Asymmetrical structures tend to be formed 
when dust grains become trapped in a high gas density vortex potentially promoting rapid growth to the planetesimal 
scale \citep{Birnstiel2013}. Dust traps are, therefore, exciting features that could be related to protoplanets 
buried in the disk. However, the detection of dust traps remains difficult in molecular lines. 

The Herbig Ae star AB Aurigae (AB Aur) hosts a well-known transitional disk. 
AB Aur has a spectral type A0-A1 \citep{Hernandez2004}. It has M$_\star$~$\sim$~2.4~M$_{{\odot}}$, T$_{\rm eff}$~$\sim$~9500~K, 
and it is located at a distance of 145 pc \citep{Ancker1998}. 
Several single-dish and interferometric molecular studies have been carried out towards
this disk \citep{Semenov2005,Corder2005,Pietu2005,Lin2006,Schreyer2008,Tang2012}.
Interferometric images of the CO J$=$2$\rightarrow$1 
and $^{13}$CO J$=$2$\rightarrow$1 lines 
and the continuum emission at 1.3 mm revealed that the AB Aur dusty disk is truncated at an inner radius of 
about 70 AU ($\sim$0$''$.5) \citep{Pietu2005}. 
Subsequent higher angular resolution observations proved the existence of a compact inner disk with a different inclination 
from that of the outer ring \citep{Tang2012}. 
The 1.3mm continuum emission from the dusty ring is highly asymmetric 
in azimuth presenting a lopsided morphology with the maximum toward the southwest. 
Thus far, this horseshoe morphology has been observed in a few transitional 
disks (e.g., IRS 48: \citealp{Marel2013}; HD 142527: \citealp{Casassus2013}) and has been interpreted as the birthsite of future planets.
This horseshoe morphology is expected when a so-called Rossby-Wave 
instability is developed at the sharp edge of the gap. The dust emission maximum corresponds to a gas pressure 
maximum where the dust particles are trapped for long timescales, a few 0.1 Myr, permitting the formation of planetesimals (km-sized particles). 

Earlier observational studies have shown a scarcity of molecular line detections
in AB Aur \citep{Semenov2005,Schreyer2008}. A similar result was found by \citet{Oberg2011}
in a larger sample of Herbig Ae stars. This is mainly due to the low molecular abundances in the
gas disk caused by the intense UV radiation from the central star 
and the freeze out of the molecules onto dust grains. 
Recently, \citet{Pacheco2015} have carried out a molecular search towards AB Aur using the IRAM
30m telescope. As a result, they detected several lines of CO and its isotopologues the HCO$^+$,
H$_2$CO, HCN, CN, and CS lines.
In addition, they detected the SO J=5$_4$$\rightarrow$3$_3$ and J=5$_6$$\rightarrow$4$_5$ lines  for the first time. AB Aur is the only protoplanetary disk detected in SO thus far, and its
detection is consistent with the interpretation of this disk being younger than
those associated with T Tauri stars.

In this paper we present high angular resolution interferometric images of the 1mm continuum, the $^{13}$CO J$=$2$\rightarrow$1, C$^{18}$O J$=$2$\rightarrow$1, H$_2$CO J$=$3$_{0,3}$ $\rightarrow$ 2$_{0,2}$, and SO J$=$5$_6$ $\rightarrow$ 4$_5$ lines.
Sulfur monoxide (SO) has been imaged in the 
transitional disk around the Herbig Ae star, AB Auriga. This species presents 
an odd spatial distribution with lack of emission toward the dust trap. 
Time-dependent chemical models are used to explain the observed behavior.

Our high angular resolutions images show, for the first time, a potential chemical 
footprint of a dust trap in a transitional disk.

\begin{figure*}
\centering
\includegraphics[width=0.9 \textwidth]{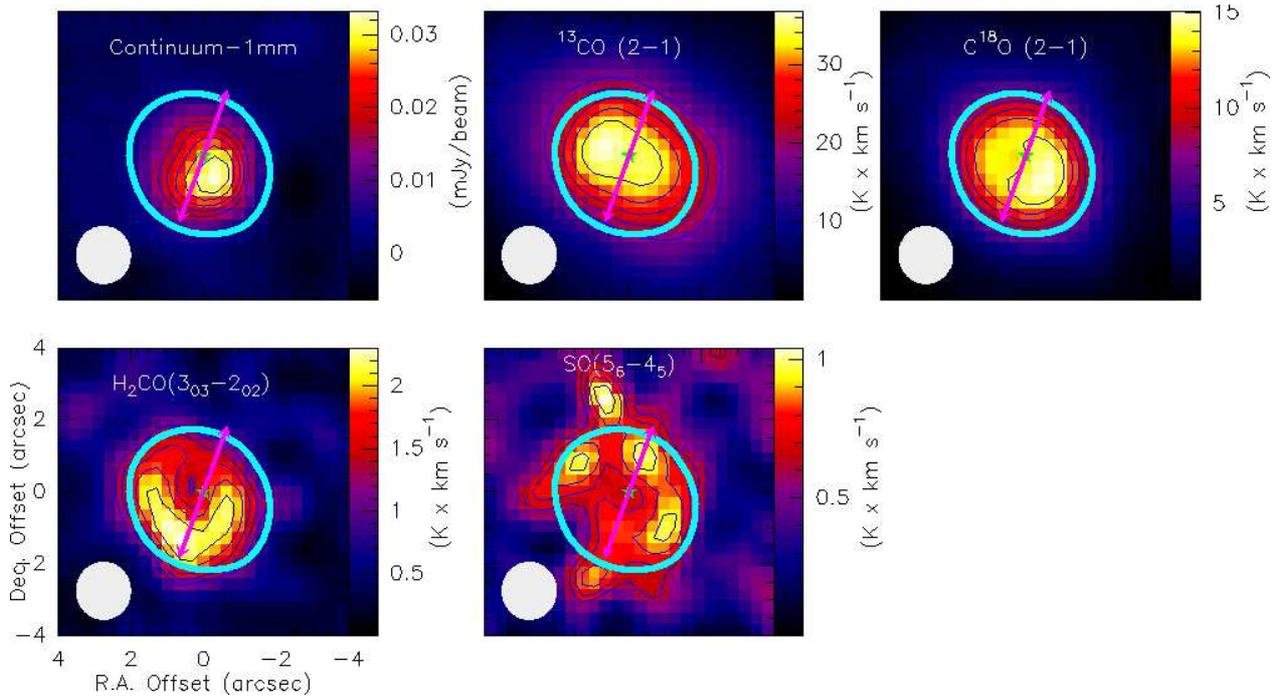}
\caption{NOEMA images of the circumstellar disk around AB Aur. The 1mm continuum dust emission map, and the
velocity-integrated intensity images from 3.5 to 8 km s$^{-1}$ of the $^{13}$CO J$=$2 $\rightarrow$1, C$^{18}$O J$=$2 $\rightarrow$1,
SO J$=$5$_6$ $\rightarrow$4$_5$, and H$_2$CO J$=$3$_{03}$ $\rightarrow$2$_{02}$ lines. White ellipses in the bottom left corner of each panel
represents the beam size ($\sim$1$''$.6$\times$1$''$.5). Contour levels are 50\% to 90\% in steps 10\% of the maximum value (34 mJy beam$^{-1}$ for the continuum, 37 K $\times$ km s$^{-1}$ for the $^{13}$CO J$=$2$\rightarrow$0, 15 K $\times$ km s$^{-1}$ for the C$^{18}$O J$=$2$\rightarrow$0, 2.3 K $\times$ km s$^{-1}$ 
for the H$_2$CO J$=$3$_{0,3}$ $\rightarrow$ 2$_{0,2}$, and 1.0 K $\times$ km s$^{-1}$ for the SO). The thick blue contour indicates the 50\% level of the C$^{18}$O emission. The arrow shows the direction of the ionized jet \citep{Rodriguez2014}.}
\label{fig:inte-maps}
\end{figure*}
\section{Observations} \label{sect:Observations}

\begin{figure}[b!]
\centering
\includegraphics[width=0.45 \textwidth]{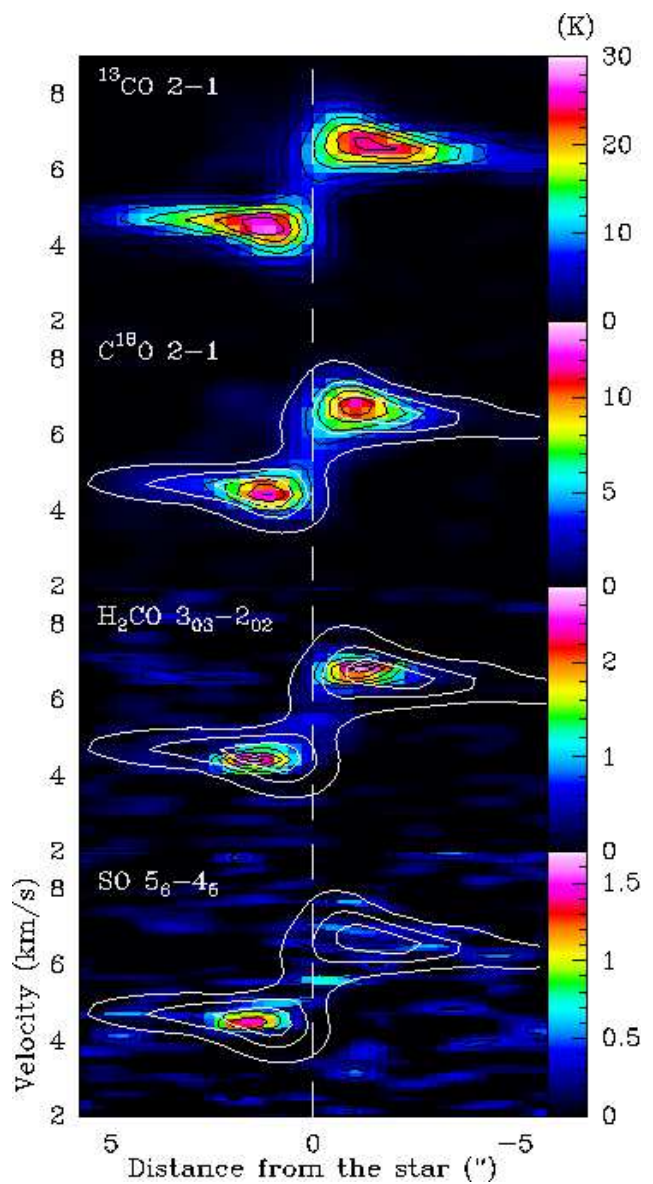}
\caption{Position-velocity diagrams along the major axis of the disk, from the NE to the SW as defined in \cite{Pietu2005}. For comparison, we have overlaid 
the contours of the $^{13}$CO J$=$2$\rightarrow$1 line in all panels. We note that there is an important asymmetry in the SO emission with very weak emission in the 
southwestern half.}
\label{fig:pv}
\end{figure}

The interferometric observations were carried out with the NOEMA interferometer in CD configuration between October and December, 2014 with six antennas providing the angular resolution and 
rms shown in Table \ref{tab:PdBI}.
We targeted AB Aur ($\alpha_{\rm J2000}$=04$\mathrm{h}$\,55\,$\mathrm{m}$ 45.8\,$\mathrm{s}$,\,$\delta_{\rm J2000}$=30$^{\circ}$ 33' 04''.2) to simultaneously observe the H$_2$CO J$=$3$_{0,3}$ $\rightarrow$2$_{0,2}$ line at 218.222 GHz, the SO J$=$5$_6$ $\rightarrow$4$_5$ line at 219.949 GHz, the C$^{18}$O J$=$2$\rightarrow$1 line at 219.560 GHz, and 
the $^{13}$CO J$=$2$\rightarrow$1 line at 220.398 GHz. The beam size was 1$''$.6$\times$1$''$.5. The channels 
free of line emission  were used to 
estimate the continuum flux that was subtracted from the spectral maps. To improve the S/N, all maps were created with a velocity  resolution of 0.25 km s$^{-1}$. 

We used 3C84, 3C454.3, MWC349, J0433, and J0418 as phase and flux calibrators. The uncertainty in the calibration is about 10\%. Data reduction and image synthesis were carried out using the GILDAS software. 

In Fig.~\ref{fig:comparacion-30m} we compare the interferometric fluxes measured with
NOEMA with those obtained with the 30m telescope by \citet{Pacheco2015}. The interferometric spectra were
obtained by fitting an elliptical Gaussian to the emission of each channel in the 
uv-plane. The interferometer recovers almost all the flux for 
the C$^{18}$O J=2$\rightarrow$1, H$_2$CO J=4$_{0,3}$$\rightarrow$2$_{0,2}$,
and SO J=5$_6$$\rightarrow4_5$ lines. As expected, a significant fraction
of the flux ($>$75\%) is missing in the $^{13}$CO J=2$\rightarrow$1 line.
We note that in some channels of the C$^{18}$O line, the NOEMA flux is higher
than that of the 30m telescope. This is physically unacceptable and most likely 
due to errors in the baseline of the 30m data. We note that the 30m observations
were performed using the wobbler switching procedure and the cloud emission is,
at least partially, subtracted with the OFF position. The absence of missing flux 
does not imply the absence of an extended component, which indeed exists
in the case of C$^{18}$O (see \citealp{Fuente2002,Fuente2010,Semenov2005}).
\section{Results} \label{sect:Results}
\subsection{NOEMA data}

Figure~\ref{fig:inte-maps} shows our interferometric images of the $^{13}$CO J$=$2$\rightarrow$1, C$^{18}$O J$=$2$\rightarrow$1, H$_2$CO J$=$3$_{0,3}$ $\rightarrow$ 2$_{0,2}$, and 
SO J$=$5$_6$ $\rightarrow$ 4$_5$ lines as observed with NOEMA (IRAM) providing an angular resolution of $\sim$1$''$.6 ($\sim$ 230 AU). The emission of the molecular lines 
coincides with the ring detected in the dust continuum emission, but significant 
differences exist among their azimuthal spatial distributions (see Fig.~\ref{fig:azimuth}). The asymmetry observed in the continuum emission is smeared out in the map of the $^{13}$CO J$=$2$\rightarrow$1 line. This is not 
surprising since this line is optically thick and it mainly traces the gas temperature at an intermediate gas layer between the disk surface and the equatorial plane. 

The spatial distribution of the C$^{18}$O J$=$2$\rightarrow$1 line is very similar to that of the 
dust continuum emission. The ratio between 1mm continuum
and C$^{18}$O J$=$2$\rightarrow$1 emission (1mm continuum)/(C$^{18}$O 2$\rightarrow$1) is uniform in the disk within a factor of $<$ 2. 
Taking into account the uncertainties in the dust temperature 
and dust opacities and based on our observations, we considered that there is no clear 
evidence that the gas-to-dust mass ratio varies inside the dust trap. We cannot discard possible variations of the gas-to-dust mass ratio at smaller scales ($<<$ 230 AU) that are not detected with the angular resolution  of our observations.
The H$_2$CO J$=$3$_{0,3}$ $\rightarrow$ 2$_{0,2}$ emission peak is not coincident with the dust trap defined by 
the 1mm continuum 
and the C$^{18}$O J$=$2$\rightarrow$1 emissions, but extends toward the south. The most extreme case is that of 
the SO J$=$5$_6$ $\rightarrow$ 4$_5$ line that presents an almost uniform emission along the ring, with an enhancement toward 
the northeast. 

The SO molecule is a well-known tracer of shocked gas and has been widely studied
in the bipolar outflows associated with low mass Class 0 and I protostars \citep{Chernin1993,Bachiller2001,
Wakelam2005,Codella2005,Tafalla2010}. 
Recently, the high spatial resolution observations provided by ALMA have permitted 
the detection of a new component
of the SO emission in very young protostars. \cite{Sakai2014}  and \cite{Podio2015} have
detected disk-size rotating rings around the Class 0 protostars L1527 and HH 212, respectively. In both cases,
they have been interpreted as the consequence of the shocks produced by the accretion of material in the 
disk-envelope interface.
In order to investigate the origin of the SO emission in AB Aur, 
we studied the velocity structure of its emission.
The velocity structure of the SO emission is consistent with the rotating disk pattern traced by $^{13}$CO J$=$2$\rightarrow$1 (see Fig. \ref{fig:pv}), suggesting that its emission could be coming from the protoplanetary disk. However, the poor S/N ratio of our observations prevents us from drawing a firm conclusion. We cannot discard other interpretations such as the emission coming from a rotating ring. Surprisingly, towards the dust trap 
the SO emission is not detected down to our sensitivity limit (rms=2 mJy/beam=0.2 K in a channel of 0.25 km~s$^{-1}$).

\begin{figure}[t!]
\centering
\includegraphics[scale=0.42]{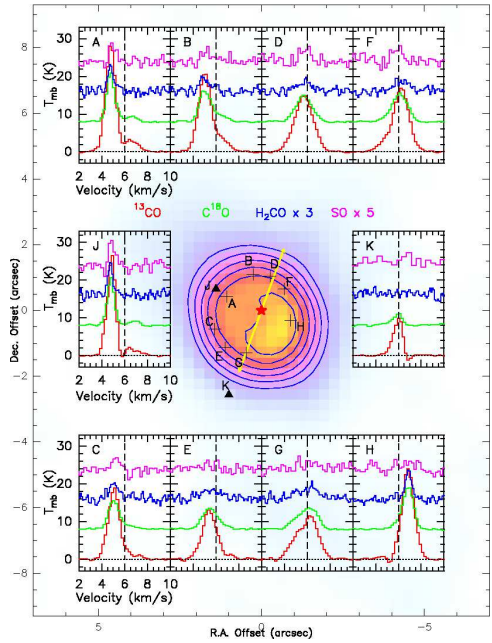}
\caption{Spectra of detected molecules over ten points around the AB Aur disk. In color gradient, the C$^{18}$O J=2$\rightarrow$1 integrated intensity map from Fig.~\ref{fig:inte-maps}.
Contour levels are from 40\% to 90\% in steps of 10\% the maximum (15 K $\times$ km s$^{-1}$). The position of the central source is marked by a red star.
The arrow indicates the direction of the ionized jet \citep{Rodriguez2014}. 
The ten boxes labeled from A to K show the spectra of 
the $^{13}$CO J$=$2$\rightarrow$1, C$^{18}$O J$=$2$\rightarrow$1, SO J$=$5$_6$$\rightarrow$4$_5$, and H$_2$CO J$=$3$_{0,3}$$\rightarrow$2$_{0,2}$ 
lines toward selected positions. We note that the SO emission is extremely weak (even undetected) toward the southwest.}
\label{fig:posiciones}
\end{figure}

Figure~\ref{fig:posiciones} shows the $^{13}$CO J$=$2$\rightarrow$1, C$^{18}$O J$=$2$\rightarrow$1, SO J$=$5$_6$$\rightarrow$4$_5$, and H$_2$CO J$=$3$_{0,3}$$\rightarrow$2$_{0,2}$ spectra from ten points labeled from A to K, overplotted with the C$^{18}$O J=2$\rightarrow$1 integrated intensity map. We note that the SO emission is extremely weak (even undetected) toward the southwest.

\begin{figure}[t!]
\centering
\includegraphics[width=0.45 \textwidth]{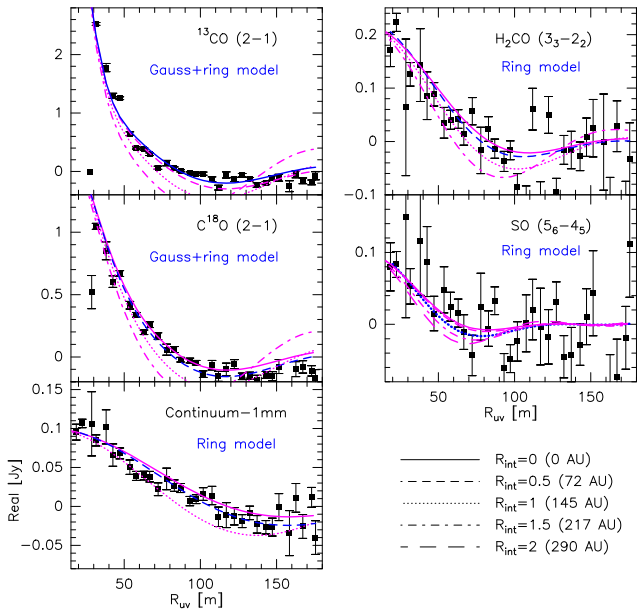}
\caption{Real part of the observed visibilities as functions of uv distance of the dust emission and the $^{13}$CO J=2$\rightarrow$1, C$^{18}$O J=2$\rightarrow$1, SO J=5$_6$$\rightarrow$4$_5$, H$_2$CO J=3$_{03}$$\rightarrow$2$_{02}$ lines. {\em Blue lines} represent the best fit models given in Table~\ref{tab:uvmodels}. {\em Magenta lines} represent the same models, but varying the inner radius from 0 to 2 arcsec in steps of 0.5.}
\label{fig:real}
\end{figure}

\subsection{Model fits in the UV plane}

\begin{table}
\centering
\caption{Fit model parameters. These values were directly obtained by fitting the visibilities.}
\begin{tabular}{llrl}  
\hline\hline
&\multicolumn{3}{c}{$^{13}$CO, rms= 1.65 Jy}\\\hline
Ring & R.A.		 & 0.0	& fixed\\
 & Dec.		 & 0.0	& fixed\\
+ & Flux		 & 1.5 	& $\pm$ 0.05 \\
 & R$_{\rm in}$ (AU)   & 0.0& $\pm$ 72 \\
 & R$_{\rm out}$ (AU)  & 286& $\pm$ 3  \\
Gaussian& R.A.		 & -0.4 & $\pm$ 0.01 \\
& Dec.		 & -0.2 & $\pm$ 0.02 \\
& Flux		 & 7.3	& $\pm$ 0.15 \\
& H.W.H.P.	 & 460  & $\pm$ 6 \\
\hline\hline
&\multicolumn{3}{c}{C$^{18}$O, rms= 1.37 Jy}\\\hline
Ring & R.A.		 & 0.0	& fixed \\
 & Dec.		 & 0.0	& fixed \\
+ & Flux		 & 0.8	& $\pm$ 0.07  \\
 & R$_{\rm in}$ (AU)   & 72 & $\pm$ 72 \\
 & R$_{\rm out}$ (AU)  & 283& $\pm$ 4     \\
Gaussian& R.A.		 & -0.05& $\pm$ 0.04  \\
& Dec.		 & -0.29& $\pm$ 0.07  \\
& Flux		 & 1.25 & $\pm$ 0.06  \\
& H.W.H.P.	 & 310  & $\pm$ 15    \\
\hline\hline
&\multicolumn{3}{c}{Continuum 1mm, rms= 0.41 Jy}\\\hline
Ring & R.A.		 &0.01 & $\pm$ 0.03 \\
 & Dec.		 &-0.30 & $\pm$ 0.04 \\
 & Flux		 &0.1   & $\pm$ 0.004 \\
 & R$_{\rm in}$ (AU)   &72  & $\pm$ 36  \\
 & R$_{\rm out}$ (AU)  &209   & $\pm$ 6 \\  
\hline\hline
&\multicolumn{3}{c}{H$_2$CO, rms=1.25 Jy}  \\\hline
Ring & R.A.		 & 0.3 & $\pm$ 0.08  \\
 & Dec.		 & -0.4& $\pm$ 0.10  \\
 & Flux		 & 0.22& $\pm$ 0.01  \\
 & R$_{\rm in}$ (AU)   & 72& $\pm$ 72  \\
 & R$_{\rm out}$ (AU)  & 302 & $\pm$ 11 \\
\hline\hline
&\multicolumn{3}{c}{SO, rms= 1.34 Jy}        \\\hline
Ring & R.A.		 & 0.5& $\pm$ 0.2  \\
 & Dec.		 & -0.3& $\pm$ 0.2  \\
 & Flux		 & 0.10& $\pm$ 0.01  \\
 & R$_{\rm in}$ (AU)   & 145 &$\pm$ 72 \\
 & R$_{\rm out}$ (AU)  & 384 & $\pm$ 38  \\\hline
\end{tabular}
\label{tab:uvmodels}
\end{table}

We have attempted to model the continuum and line emissions in 
order to have a more accurate estimate of the dust and gas spatial distributions.
Fig.~\ref{fig:real} shows the real part of the visibilities as a function of the radius 
in the uv plane. In the case of $^{13}$CO and C$^{18}$O, the curve does not cut 
the y-axis, which indicates the existence of an extended component whose flux is not 
fully recovered by our observations.
Fig.~\ref{fig:imaginary} shows the imaginary part of the visibilities as a function of the radius.
Values of the imaginary part different from zero indicate an azimuthal 
asymmetry \citep{Pinilla2015c}. The azimuthal asymmetry is clear in $^{13}$CO, C$^{18}$O, and the 
continuum emissions. In the case of $^{13}$CO, the asymmetry is located at very short uv distances (i.e., large radii) and 
is more likely related to the extended emission component.
The plots of H$_2$CO and SO suggest more symmetric distributions.  
We used a ring model to fit the continuum, H$_2$CO, and SO data and 
we used a ring + circular Gaussian model to fit $^{13}$CO and C$^{18}$O. 
The best fits to these data are shown in Table~\ref{tab:uvmodels} and the residual images in 
Fig.~\ref{fig:mapas-residuos}. For $^{13}$CO and C$^{18}$O, we have significant residuals towards the east, 
because there is extended emission coming from a flattened envelope. The continuum, H$_2$CO, and SO data are 
well fitted with a  ring model. As expected, there is some weak  
continuum and C$^{18}$O residual emissions towards the dust trap. 
Our fits prove that the emission of the continuum is more compact than that of the molecular lines.
Moreover, the SO emission has larger inner and outer radii than the other molecules, suggesting that 
it is coming from an outer part of the disk. The width of the ring, $\sim$200~AU, is 
significantly larger than those detected in the case of the younger objects 
L1527 and HH212 \citep{Sakai2014,Podio2015}.

\section{Abundance calculations}\label{sect:abundance_calculations}

\begin{figure}[b!]
\centering
\includegraphics[width=0.47 \textwidth]{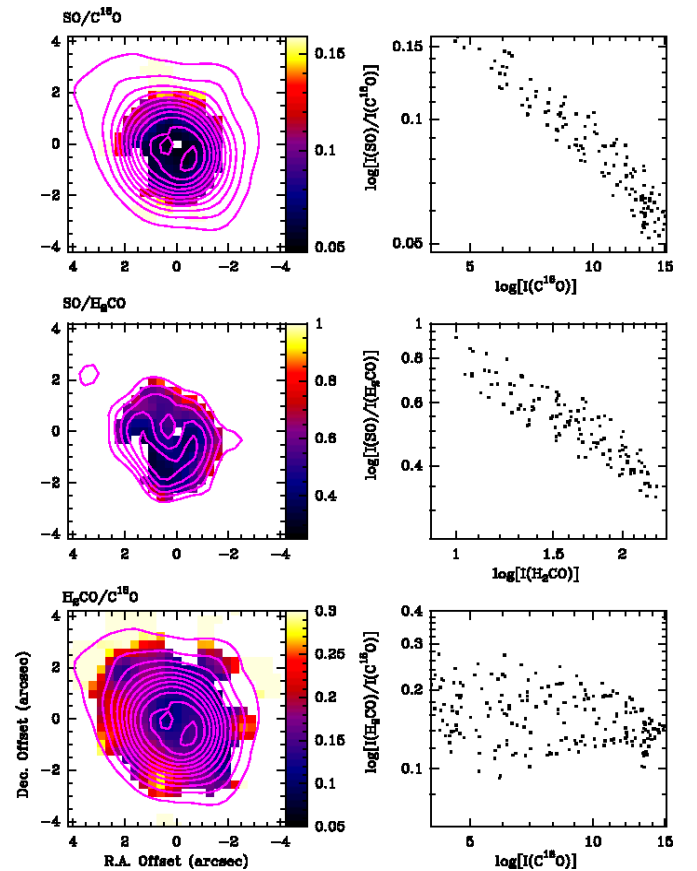}
\caption{Line integrated intensity ratios. The color maps correspond to 
I(SO J$=$5$_6$$\rightarrow$4$_5$)/I(C$^{18}$O  J$=$2$\rightarrow$1), I(SO J$=$5$_6$$\rightarrow$4$_5$)/I(H$_2$CO J$=$3$_{0,3}$$\rightarrow$2$_{0,2}$), and
I(H$_2$CO  J$=$3$_{03}$$\rightarrow$2$_{02}$)/I(C$^{18}$O J$=$2$\rightarrow$1) in temperature units. While the I(H$_2$CO)/I(C$^{18}$O) ratio 
is fairly constant in the disk, the I(SO)/I(C$^{18}$O) and I(SO)/I(H$_2$CO) present a clear NE-SW gradient with its minimum value toward the 
dust trap. Although there is a large dispersion, there appears to be an anticorrelation with C$^{18}$O intensity, which is a measure of 
column density, which in turn is thought to correlate (assuming a disk-like or ring morphology) with the local density.}
\label{fig:nubes}
\end{figure}

To investigate the chemical structure of the AB Aur disk, we produced maps of the line integrated intensity ratios of the observed SO, H$_2$CO, and C$^{18}$O lines (see Fig. \ref{fig:nubes}).  In the case of optically thin emission and uniform physical conditions along the disk, these maps would be proxy of the relative abundances. The I(SO J$=$5$_6$$\rightarrow$4$_5$)/I(C$^{18}$O  J$=$2$\rightarrow$1) and I(SO J$=$5$_6$$\rightarrow$4$_5$)/I(H$_2$CO J$=$3$_{0,3}$$\rightarrow$2$_{0,2}$) line integrated intensity ratios present a clear northeast-southwest gradient  (a factor of 3$-$6) with the minimum in the dust trap as defined by the continuum peak at 1mm. The I(H$_2$CO  J$=$3$_{03}$$\rightarrow$2$_{02}$)/I(C$^{18}$O J$=$2$\rightarrow$1) ratio is fairly constant, with values 
from $\sim$0.15 in the western part to $\sim$0.2 in the eastern part of the disk, but towards the south this ratio increases to values $>$ 0.3, which are a factor of 2 higher than the averaged value in the disk.
This position is coincident with the jet direction, which means that the enhancement of the H$_2$CO abundance might be related to the jet-disk interaction.

One possibility to explain the weak SO emission towards the dust trap would be that the envelope or larger scale emission are optically thick and mask the emission at certain velocities 
and thus, at certain positions in the disk. We detect the weak SO emission towards the dust trap. The emission from the dust trap 
is at $\sim$7~km~s$^{-1}$ (see Fig.~\ref{fig:pv}), i.e., about 1~km~s$^{-1}$ shifted from the central velocity of the molecular cloud 
\citep{Fuente2010}. The emission from the cloud is narrow, $\Delta v$ = 0.5~km~s$^{-1}$, and centered at $v$ = 5.9~km~s$^{-1}$,
making it impossible to mask the emission from the dust trap.

\begin{figure}[b!]
\centering
\includegraphics[width=0.45 \textwidth]{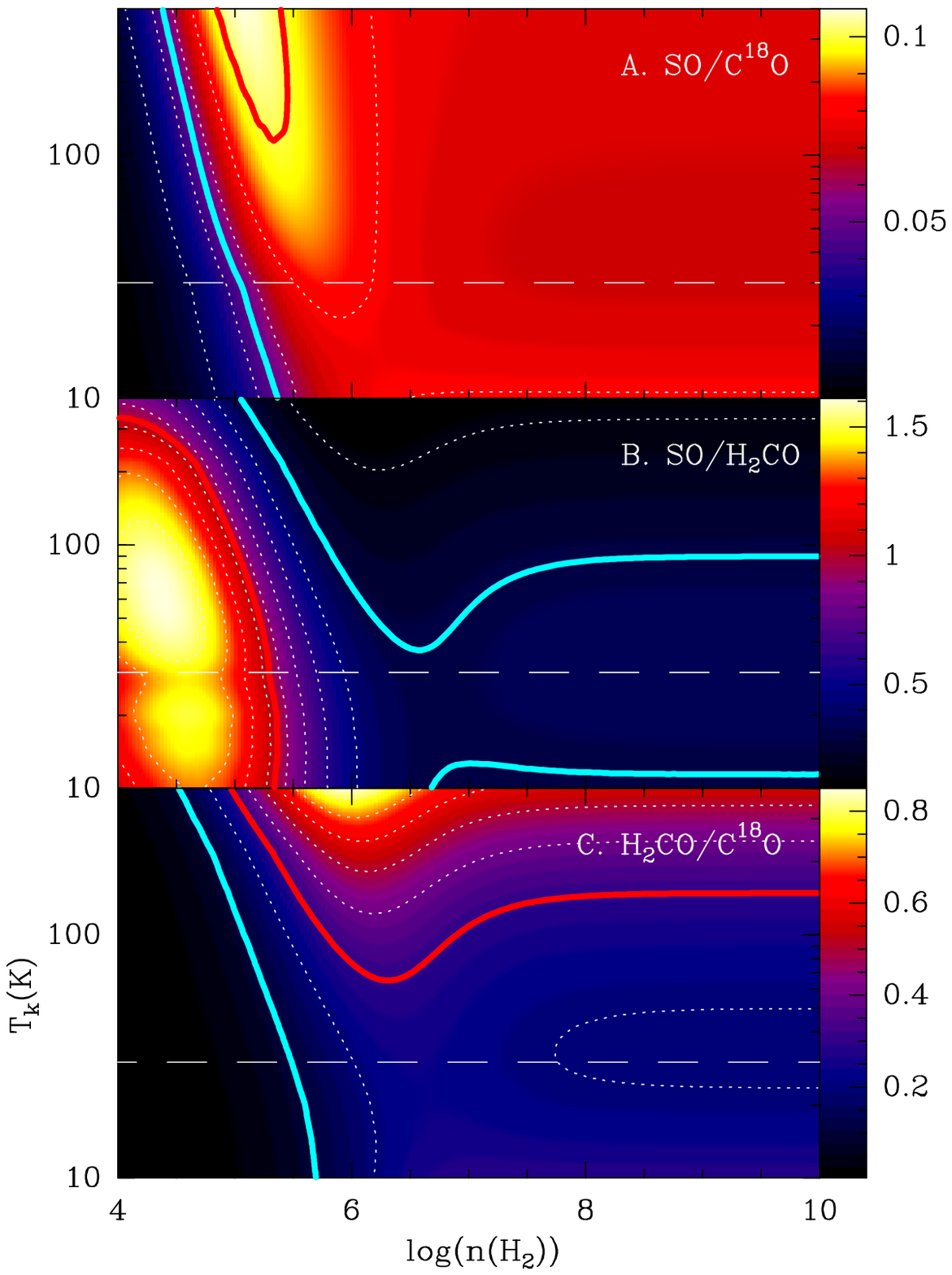}
\caption{LVG calculations. We assumed N(SO)=2$\times$10$^{13}$~cm$^{-2}$, 
N(H$_2$CO)=1$\times$10$^{13}$~cm$^{-2}$, and N(C$^{18}$O)= 2$\times$10$^{16}$~cm$^{-2}$.
From top to bottom, we plot the I(SO J$=$5$_6$$\rightarrow$4$_5$)/I(C$^{18}$O  J$=$2$\rightarrow$1), 
I(SO J$=$5$_6$$\rightarrow$4$_5$)/I(H$_2$CO J$=$3$_{0,3}$$\rightarrow$2$_{0,2}$), 
and I(H$_2$CO  J$=$3$_{03}$$\rightarrow$2$_{02}$)/I(C$^{18}$O J$=$2$\rightarrow$1) line intensity ratios. The {\em long-dashed} line corresponds to a gas kinetic temperature of T$\rm_k$=30 K, 
which is the midplane temperature 
derived by \cite{Pietu2005} and \cite{Pacheco2015}. {\em Solid lines} indicate the values derived from our maps (Fig. \ref{fig:nubes}).
The red 
and blue lines trace the maximum and minimum observed ratios, respectively.}
\label{fig:density}
\end{figure}

LVG (Large Velocity Gradient) calculations have been carried out to determine whether the gradients observed in the I(SO J$=$5$_6$$\rightarrow$4$_5$)/I(C$^{18}$O  J$=$2$\rightarrow$1) and I(SO J$=$5$_6$$\rightarrow$4$_5$)/I(H$_2$CO J$=$3$_{0,3}$$\rightarrow$2$_{0,2}$) ratios are due to the different physical conditions (higher densities and slightly lower temperatures in the dust trap) or to 
variations in the chemical composition of the gaseous disk. Fig. \ref{fig:density} shows the results of our LVG calculations for a grid of reasonable physical conditions in the disk. The molecular column densities have been 
fixed to N(SO)=2$\times$10$^{13}$ cm$^{-2}$, N(H$_2$CO)=1$\times$10$^{13}$ cm$^{-2}$, and N(C$^{18}$O)= 2$\times$10$^{16}$ cm$^{-2}$. The velocity dispersion is assumed to be 1.5 km s$^{-1}$ and the gas kinetic temperature and molecular 
hydrogen density vary within the range of values, T$\rm_k$ = 20 – 510 K and n(H$_2$)= 10$^4$ – 10$^{10}$ cm$^{-3}$.  

In Fig. \ref{fig:density}, we plot the calculated I(SO J$=$5$_6$$\rightarrow$4$_5$)/I(C$^{18}$O  J$=$2$\rightarrow$1), 
I(SO J$=$5$_6$$\rightarrow$4$_5$)/I(H$_2$CO J$=$3$_{0,3}$$\rightarrow$2$_{0,2}$), and   
I(H$_2$CO  J$=$3$_{03}$$\rightarrow$2$_{02}$)/I(C$^{18}$O J$=$2$\rightarrow$1) ratios as a function of gas kinetic 
temperature and molecular hydrogen density. The long-dashed line corresponds to a gas kinetic temperature of T$\rm_k$=30 K, which is the midplane temperature as 
derived by \cite{Pietu2005} and \cite{Pacheco2015}. Solid lines indicate the values derived from our maps (Fig. \ref{fig:nubes}). 
The gas kinetic temperature does not have a major effect in the calculated 
line intensity ratios unless we go to unrealistic values of T$\rm_k$ $>$ 100~K. 
The observed species are easily photodissociated and their emissions are not expected to come from
the highly irradiated surface but from colder regions in between the surface and the midplane.

The values of 
I(SO J$=$5$_6$$\rightarrow$4$_5$)/I(C$^{18}$O  J$=$2$\rightarrow$1), 
I(SO J$=$5$_6$$\rightarrow$4$_5$)/I(H$_2$CO J$=$3$_{0,3}$$\rightarrow$2$_{0,2}$), and 
I(H$_2$CO  J$=$3$_{03}$$\rightarrow$2$_{02}$)/I(C$^{18}$O J$=$2$\rightarrow$1) observed in the eastern 
half can be explained with a density of a few $\times$10$^6$ cm$^{-3}$ (red lines
in Fig. \ref{fig:density}), which is consistent with most of the molecular emission 
coming from an intermediated layer between the midplane and the disk surface.
We encounter difficulties, however, when fitting the values observed towards the dust trap (blue lines in Fig. \ref{fig:density}). 
Assuming the same N(SO)/N(C$^{18}$O) and N(CO)/N(H$_2$CO) ratios as for the eastern half, 
the low values of I(SO J$=$5$_6$$\rightarrow$4$_5$)/I(C$^{18}$O  J$=$2$\rightarrow$1) towards the dust trap 
would require a hydrogen density lower than 10$^5$ cm$^{-3}$.
However, the I(SO J$=$5$_6$$\rightarrow$4$_5$)/I(H$_2$CO J$=$3$_{0,3}$$\rightarrow$2$_{0,2}$) would need gas kinetic temperatures $>$ 200 K 
and a density of $\sim$10$^6$ cm$^{-3}$. The only way to explain the observations is to decrease the SO abundance with respect 
to C$^{18}$O.
This would imply that the SO abundance is changing in the protoplanetary disk with a minimum in the dust trap. This decrease 
in the SO abundance can be considered the chemical footprint of the local pressure maximum. 

In Table \ref{tab:abundances}  we show the  molecular column densities and fractional abundances estimated for the positions shown 
in Fig. \ref{fig:posiciones}. In these calculations, 
we assume T$\rm_k$= 30 K and n(H$_2$)= 1$\times$10$^6$ cm$^{-3}$. The SO abundances were derived from the obtained
N(SO)/N(C$^{18}$O) ratios assuming a canonical C$^{18}$O abundance of X(C$^{18}$O)=1.7$\times$10$^{-7}$. 
Values of the SO abundance close to 2$\times$10$^{-10}$ are found in the eastern part, and 2-3 times lower toward the dust trap. 
This abundance is $>$ 4 orders of 
magnitude smaller than those measured in bipolar outflows \citep{Bachiller1997} and about 3 orders of magnitude lower than 
those in young disks \citep{Sakai2014, Podio2015}.
This suggests that an important evolution occurs in the chemistry of the gas between accretion disk phase and the formation of a planetary 
system, and SO is tracing non-shocked gas within the PPD around AB Aur.

Following the same procedure, we derive H$_2$CO fractional abundances of a 
few 10$^{-10 }$ in this protoplanetary disk. This abundance is a factor of 20 lower than that derived toward the IRS 48 transitional 
disk on the basis of high spatial resolution (25 - 38 AU)  observations \citep{Marel2014}. However, it is similar to the 
averaged H$_2$CO abundances found in other protoplanetary 
disks \citep{Fuente2010}. In fact, IRS 48 is the only protoplanetary disk with such a high H$_2$CO abundance thus far. 
Further observations are required to determine 
whether IRS 48 has a peculiar chemistry compared with other transitional disks.

In this section, we assume that C$^{18}$O, H$_2$CO, and SO are coeval and that their emissions 
come from a uniform layer with constant temperature and density. One would expect, however, 
the existence of structure at scales $<$200~AU. For instance, we cannot discard 
the possibility that the emission of SO and/or H$_2$CO is concentrated in several
unresolved rings, each one with a different abundance relative to C$^{18}$O. However, the results of our fittings (see Table~\ref{tab:uvmodels}) 
do not support this interpretation. We also note also that 
the abundances of these compounds are expected to vary by several orders of magnitude in the vertical scale,
from the midplane to the disk surface \citep{Pacheco2015}. Higher spatial resolution
observations and a multitransitional study are necessary to further constrain the H$_2$CO and SO spatial distributions and have a deeper insight into their chemistry.

\begin{table*}
\centering
\caption{LVG estimation for column densities and abundances. We adopted a molecular hydrogen density
of n(H$_2$)= 1$\times$10$^6$ cm$^{-3}$ and T$\rm_k$= 30 K. Fractional abundances are calculated assuming X(C$^{18}$O)=1.7$\times$10$^{-7}$.
The velocity dispersion is assumed to be 1.5 km s$^{-1}$. Notation: 1(15) means 1$\times$10$^{15}$.}
\begin{tabular}{lllllcc}       
\hline\hline
Positions &N$_{\rm C^{18}O}$ &N$_{\rm p-H_2 CO}$&N$_{\rm SO}$&\multirow{2}{*}{$\frac{N_{\rm SO}}{N_{\rm H_2}}$}&\multirow{2}{*}{$\frac{N_{\rm p-H_2CO}}{N_{\rm H_2}}$ $\times$ 4}&
\multirow{2}{*}{$\frac{N_{\rm SO}}{N_{\rm p-H_2CO}}$ $\times$0.25}\\
            &cm$^{-2}$       &   cm$^{-2}$      &cm$^{-2}$     &                                              &                            &                      \\\hline
A& 9.5(15)   &	4.0(12)  &	1.0(13)  &	1.7(-10)  &	2.9(-10)  &	0.6 \\
B& 1.0(16)   &	4.0(12)  &	7.0(12)  &	1.1(-10)  &	2.7(-10)  &	0.4 \\
C& 7.0(15)   &	5.5(12)  &	6.0(12)  &	1.4(-10)  &	5.4(-10)  &	0.3 \\
D& 1.1(16)   &	4.0(12)  &	1.0(13)  &	1.5(-10)  &	2.5(-10)  &	0.6 \\
E& 7.5(15)   &	5.0(12)  &   $<$ 6.0(12)  &   $<$ 1.4(-10)  &	4.5(-10)  &  $<$ 0.3 \\
F& 1.1(16)   &	4.5(12)  &	1.0(13)  &	1.5(-10)  &	2.8(-10)  &	0.5 \\
G& 1.1(16)   &	7.0(12)  &   $<$ 9.0(12)  &   $<$ 1.4(-10)  &	4.3(-10)  &  $<$ 0.3 \\
H& 1.2(16)   &	4.7(12)  &   $<$ 5.0(12)  &   $<$ 7.1(-11)  &	2.7(-10)  &  $<$ 0.3 \\
J& 8.0(15)   &	4.4(12)  &	1.0(13)  &	2.1(-10)  &	3.7(-10)  &	0.6 \\
K& 2.2(15)   &$<$ 4.0(12) &   $<$ 5.0(12)  &   $<$ 3.9(-10)  &  $<$ 1.2(-09)  &	- \\
       \hline         
\end{tabular}
\label{tab:abundances}
\end{table*}

\section{Chemical model}
\label{sect:chemical_models}

\begin{figure*}
\centering
\includegraphics[width=0.79 \textwidth]{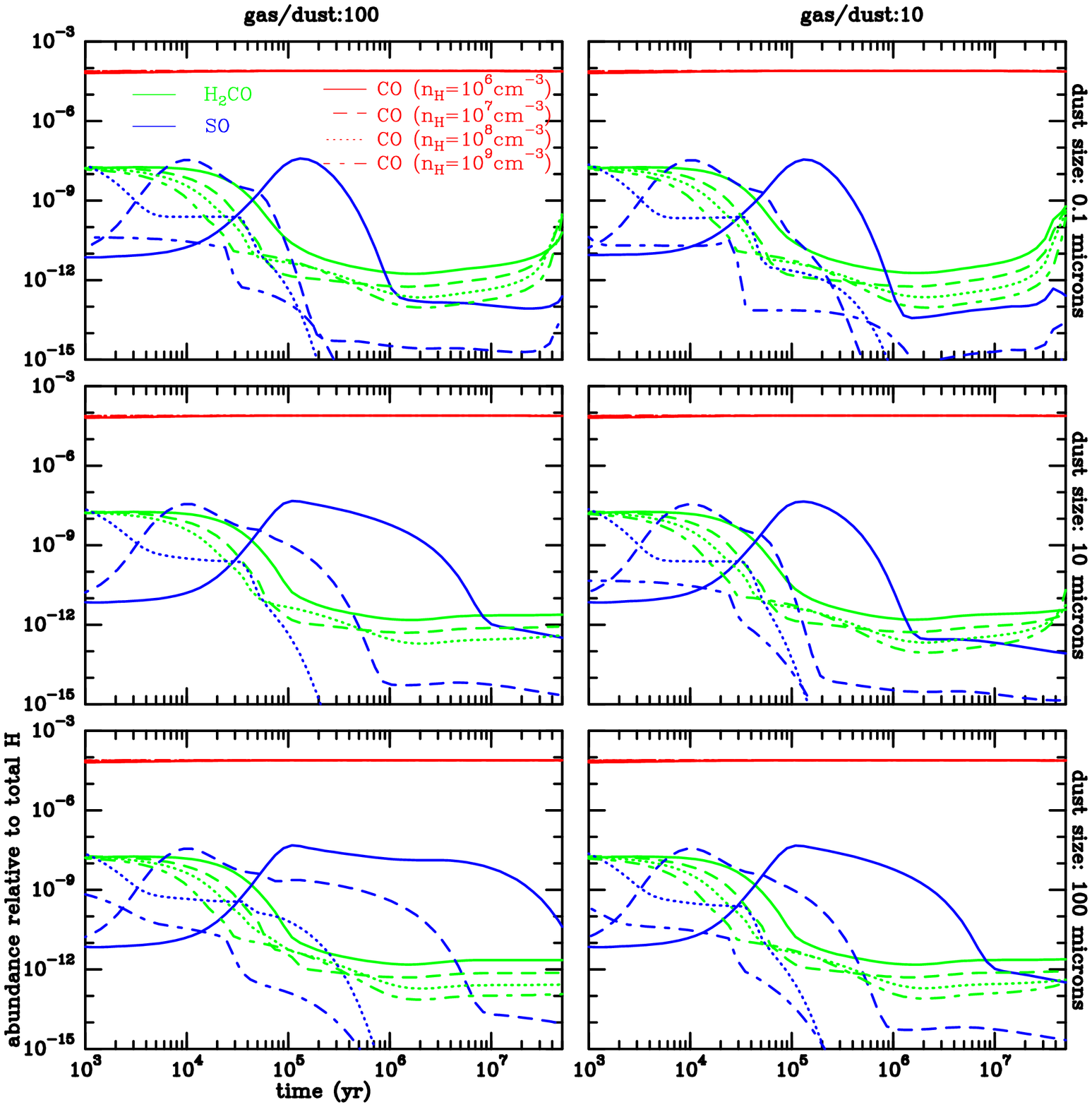}
\caption{Chemical model abundance results. Predicted CO, H$_2$CO, and SO fractional abundances as a function of time for different molecular hydrogen densities, gas-to dust mass ratios, and grain sizes.}
\label{fig:chemical-models}
\end{figure*}

The time-dependent chemical model described by \cite{Pacheco2015} was used to investigate the 
molecular chemistry in the dust trap. This model is an updated version of that reported 
by \cite{Agundez2008} and \cite{Fuente2010} and includes the elements H, C, N, O, and S. 
The model considers adsorption onto dust grains and 
desorption processes, such as thermal evaporation, photodesorption, and desorption induced by 
cosmic rays, but surface reactions are not included. 
Desorption energies are adopted from the compilations
by \citet{Hasegawa1993} and \citet{Willacy2009}. All the grains are assumed to 
be of the same size.

The initial 
molecular composition corresponds to that of a dark cloud 
(n$_{\rm H}$=2$\times$10$^4$~cm$^{-3}$, T$_k$=10~K) at a time
of 0.1~Myr assuming the so-called low metal values 
(model M1 in \citealp{Pacheco2015}). The initial ionization fraction is 7$\times$10$^{-8}$, which is typical of dark 
cores \citep{Caselli1998,Agundez2013}. 
We assume that the molecules are located close to the midplane and well shielded 
from the stellar UV radiation (A$_v$$>$10~mag). The gas and dust temperature is fixed 
to 45~K, which is slightly
higher than that derived from \citet{Pietu2005} from interferometric observations and assumed for the abundance calculations here. We note, however, that
the value of the temperature is not relevant for our results as long as T$\sim$$<$50~K, 
the SO evaporation temperature; a lower temperature would produce a larger
SO depletion. We vary the grain size, gas-to-dust mass ratio, and the molecular 
hydrogen density in order to investigate the possible effect of the dust trap on the gas chemistry.

Figure~\ref{fig:chemical-models} shows the CO, H$_2$CO, and SO abundances as a 
function of time for different molecular hydrogen densities, gas-to-dust mass ratios, 
and grain sizes. The gas-to-dust mass ratio has a minor effect on the chemistry of
these compounds. The grain size has some impact on the SO abundance.
For large grains, the SO remains in gas phase 
until later times because of the decrease in the total grain surface. In particular, 
the molecular hydrogen density is the parameter that
determines the chemistry of this species.

The SO abundance presents variations
of more than 3 orders of magnitude between 0.1 and a few Myr depending on the 
molecular hydrogen density. For high densities ($>$10$^7$ cm$^{-3}$), the SO 
abundance decreases to values $<$10$^{-12}$ in less than 0.1~Myr, the typical age
of the dust trap.  
This is due to the adsorption of SO on dust grains and the rapid conversion of SO into SO$_2$ via the reaction 
SO$+$O$\rightarrow$SO$_2$+h$\nu$. The desorption energy of SO$_2$ is higher (3070 K) than that of
SO (desorption energy= 2000 K) and once formed, is rapidly frozen onto the grain mantles. 
Below the SO evaporation temperature, the depletion of SO and SO$_2$ is proportional to the gas 
density. These predictions could explain the low average density obtained from our LVG calculations 
($<$10$^7$~cm$^{-3}$) and the uniform SO emission all over the disk.

In the case of H$_2$CO, the desorption energy
is lower, 1760 K, and adsorption is less important than
for SO, although its abundance decreases at later times because of the decrease of the atomic
carbon in gas phase that becomes locked in CO. The decrease of the H$_2$CO abundance is mainly
a consequence of the gas phase chemistry and contrary to SO, initial conditions have a large
impact on the resulting H$_2$CO abundance (see discussion by \citealp{Pacheco2015}.)

The CO abundance is practically constant over time and is equal to the 
assumed initial C elemental abundance. On the basis of these calculations, the 
differentiated CO, H$_2$CO, and SO spatial distributions mainly trace the 
gas density structure within the disk. The SO abundance is very sensitive to the gas density 
and might be a good tracer of the disk evolution.

We have to note, however, that there are large uncertainties inherent in the modeling of the sulfur chemistry. First of all, the binding energies of SO and H$_2$CO depend on the chemical nature of the substrate. \citet{Garrod2006} consider a water ice substrate and give higher desorption energies, by 15\% and 30\% for SO and H$_2$CO, respectively, than the values computed by \citet{Hasegawa1993}. Adopting these higher values we obtain the same qualitative behavior for the SO abundance, which decreases with increasing density, while the H$_2$CO abundance does not vary significantly. The decrease in the abundance of SO is related to the rapid conversion into SO$_2$ via the radiative association with atomic oxygen and the further depletion of SO$_2$ onto dust grains. More important are the uncertainties in the rate constant of the SO + O reaction and in the amount of atomic oxygen in the gas phase. The rate coefficient adopted for the reaction of SO and O, $3.2\times10^{-16}({\rm T}/300)^{-1.5}$ cm$^3$ s$^{-1}$, is based on a theoretical calculation by \citet{Millar1990}, although it has a large uncertainty as do most radiative associations. Initial conditions also play a role in our results since the amount of atomic oxygen present in the gas phase depends on the assumed elemental C/O abundance ratio, among other physical and chemical parameters.

\section{Summary and conclusions}
\label{Summary}

We present the high spatial resolution images of the  $^{13}$CO J=2 $\rightarrow$1, C$^{18}$O J=2 $\rightarrow$1, SO J=5$_6$ $\rightarrow$4$_5$, and H$_2$CO J=3$_{03}$ $\rightarrow$2$_{02}$ lines toward the transitional disk around AB Aur. Our results and conclusions can be summarized as follows.

   \begin{enumerate}
	\item The integrated intensity images of CO and H$_2$CO lines and the dust 1 mm continuum emission toward AB Aur disk trace a lopsided horseshoe-shaped distribution. Sulfur monoxide presents an odd spatial distribution with lack of emission toward the dust trap. 
	
	\item The I(SO)/I(C$^{18}$O) and I(SO)/I(H$_2$CO) ratios present a clear NE-SW gradient with its minimum value toward the dust trap. Although with a large dispersion, there is an anticorrelation of the I(SO)/I(C$^{18}$O) and I(SO)/I(H$_2$CO) ratios with the gas density. Our calculations show that the I(SO)/I(H$_2$CO) and the I(SO)/I(C$^{18}$O) ratios cannot be explained by spatial variations in the physical conditions, but by the decrease of the SO abundances toward the dust trap.
	
	\item The time-dependent model described by \cite{Pacheco2015} was used to follow the chemical evolution of the gas in the dust trap. Assuming the chemical composition of a dark cloud as the initial conditions, under high density conditions (n(H$_2$)>10$^7$ cm$^{-3}$), the SO abundance would be drastically reduced in less than 0.1 Myr. However, the CO abundance would remain constant as long as the gas temperature is >25 K.

\hspace{0.4cm} The physical conditions in the AB Aur disk are far from being uniform, with strong temperature and density gradients in the vertical scale due to the heating of the disk surface by the UV radiation from the star. The physical conditions also vary in azimuth, with the gas pressure maximum located in the southwestern part of the disk.
Because of the strong dependence of SO abundance on the gas density, we propose that the SO emission does not arise from the midplane but from an intermediate layer between the midplane and disk surface with a moderate density of a few 10$^6$~cm$^{-3}$.
For this reason, its emission seems to avoid the high pressure vortex associated with the dust trap, which is clearly detected in the dust continuum emission and C$^{18}$O.

\hspace{0.4cm} AB Aur is a good example of the tight interplay of the gas dynamics, grain growth and gas chemistry in the planet formation process.
Because of its characteristics, the sulfur chemistry is potentially an excellent tool to understanding the planet formation process and detecting the birthsites of future planets. The large uncertainties inherent
to chemical models and the scarcity of molecules detected in disks are still important drawbacks.

\hspace{0.4cm} Future ALMA and NOEMA observations could allow a full test of the
chemistry of simple and more complex molecules in transitional
disks such as those around AB Aur.\\

\end{enumerate}	

\begin{acknowledgements}
 We thank the Spanish MINECO for funding support from
 grants CSD2009-00038, FIS2012-32096 and AYA2012-32032,
 and ERC under ERC-2013-SyG, G. A. 610256 NANOCOSMOS.  
 S. P-V acknowledge the financial support of CONACyT, Mexico.
 We thank the referee for the fruitful comments.
\end{acknowledgements}

\bibliography{dust_trap}
\bibliographystyle{aa}

\clearpage
\begin{appendix}
\section{Figures}

\begin{figure}[b!]
\centering
\includegraphics[width=0.4 \textwidth]{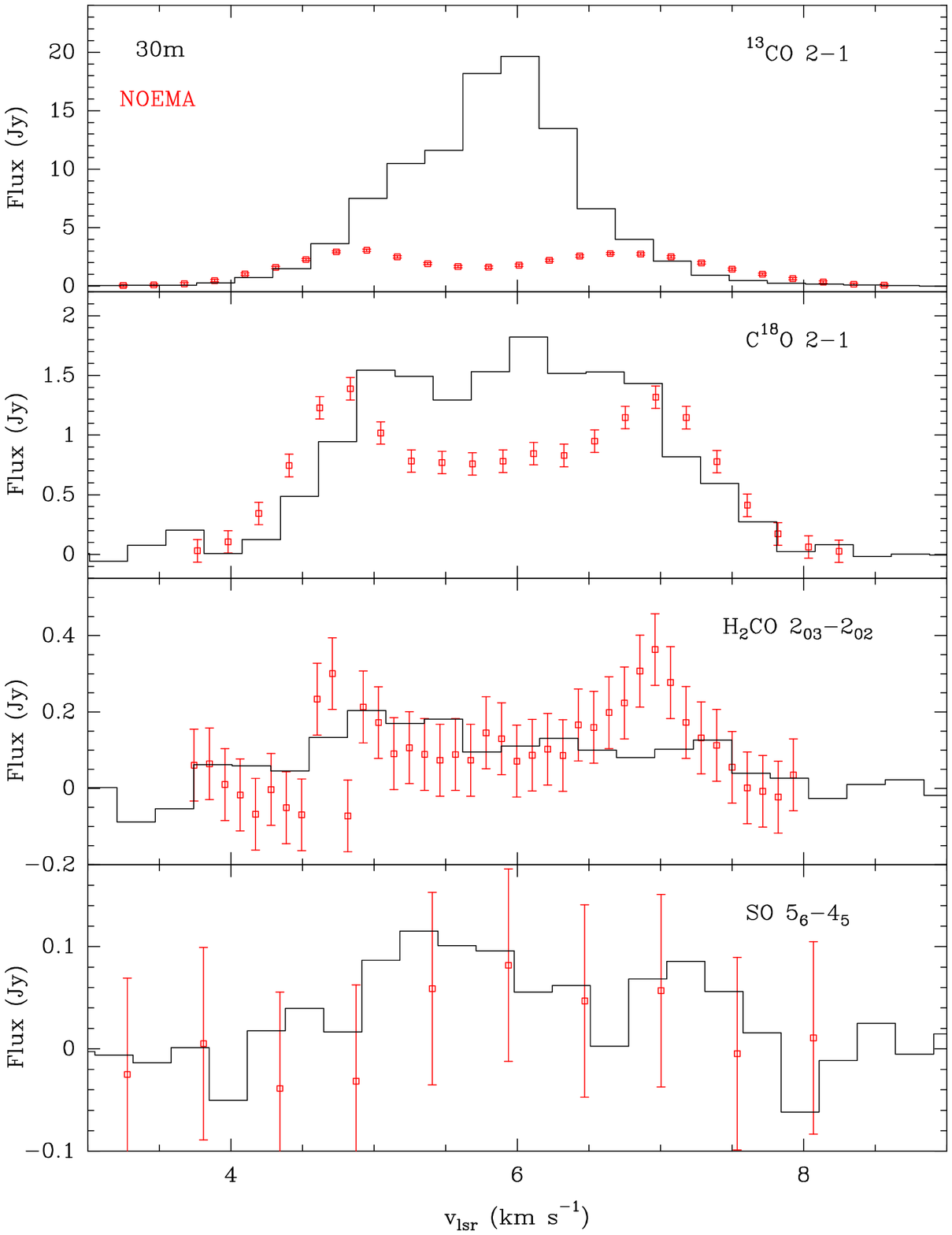}
\caption{Comparison between the single-dish IRAM 30m spectra ({\em black histogram}) 
published by \citet{Pacheco2015} and NOEMA interferometric 
spectra ({\em red points}). We note that the 30m observations
were performed using the wobbler switching procedure and the cloud emission is subtracted with the OFF position. The absence of missing flux 
does not imply the absence of an extended component.}
\label{fig:comparacion-30m}
\end{figure}

\begin{figure}[b!]
\centering
\includegraphics[width=0.5 \textwidth]{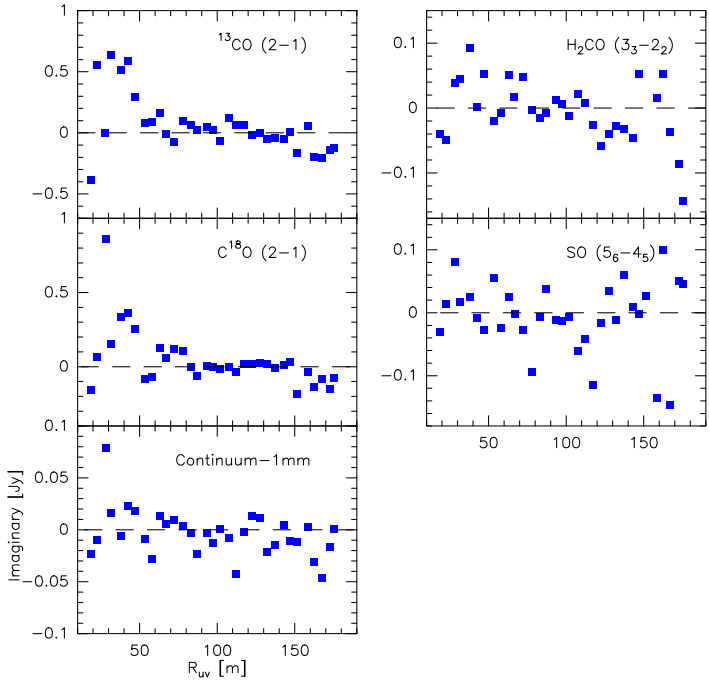}
\caption{Imaginary part of the observed visibilities for the dust emission and the $^{13}$CO J=2 $\rightarrow$1, C$^{18}$O J=2 $\rightarrow$1, SO J=5$_6$ $\rightarrow$4$_5$, and H$_2$CO J=3$_{03}$ $\rightarrow$2$_{02}$ lines.}
\label{fig:imaginary}
\end{figure}

\begin{figure*}[]
\centering
\includegraphics[width=0.85 \textwidth]{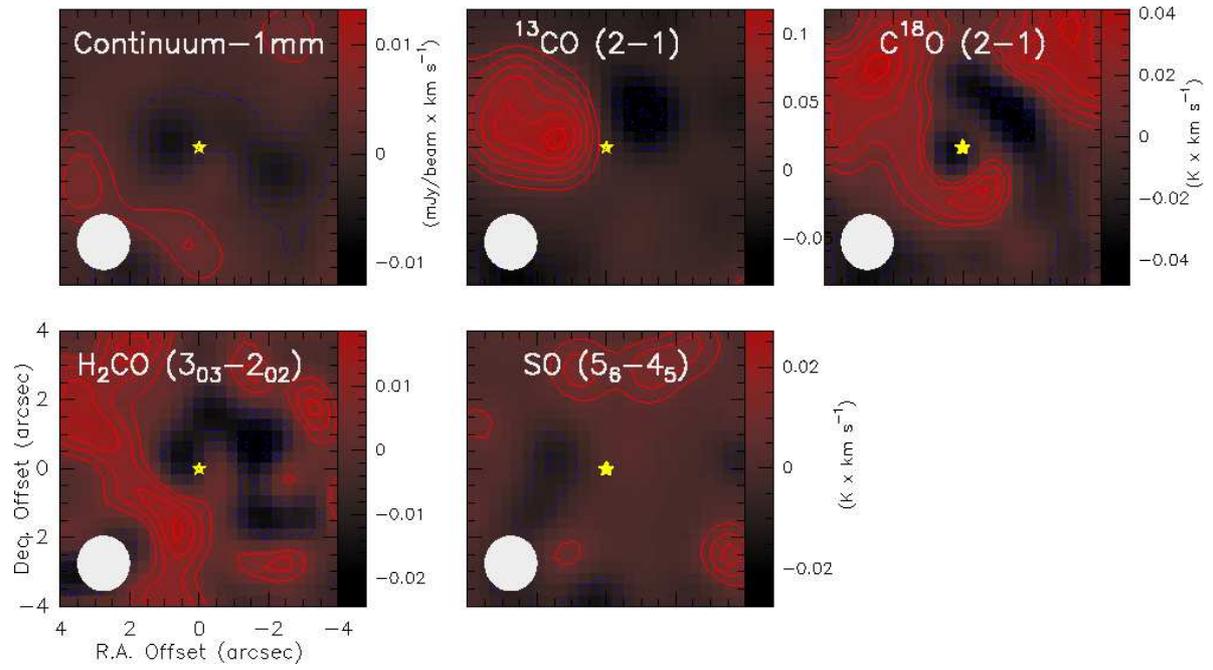}
\caption{Residual maps after subtracting the best fit model from Table \ref{tab:uvmodels}.}
\label{fig:mapas-residuos}
\end{figure*}

\begin{figure}[t!]
\centering
\includegraphics[width=0.45 \textwidth]{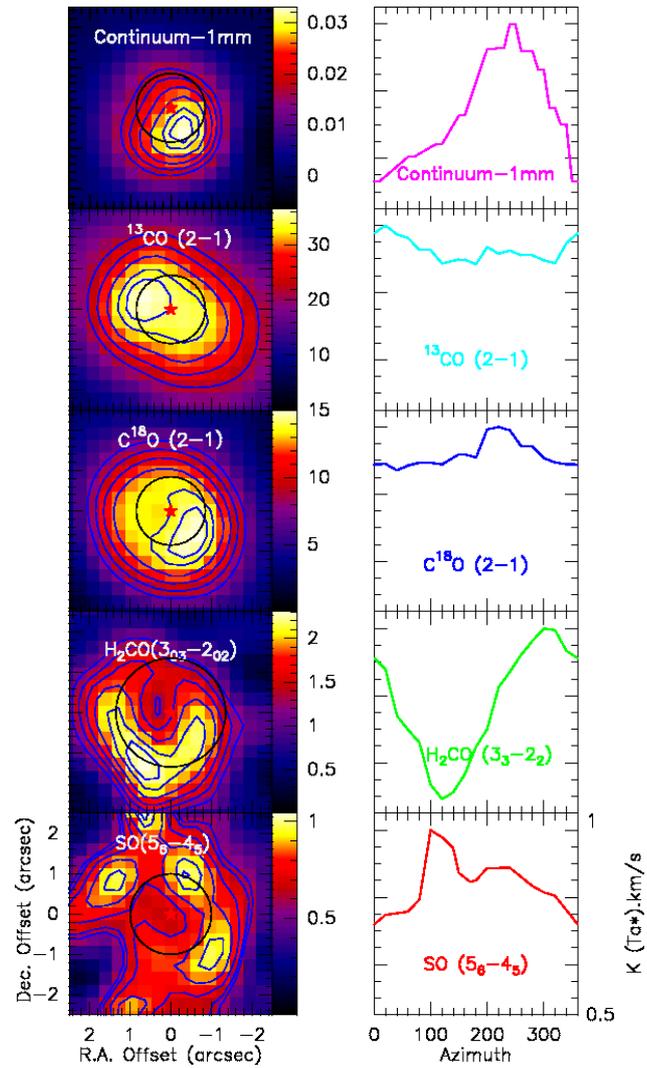}
\caption{Molecular intensity as a function of the azimuthal angle ({\em right}) over a circle of 0.8 arcsec for the continuum, the $^{13}$CO and the C$^{18}$O maps, and 1.3 and 1 arcsec for H$_2$CO and SO, respectively, around the central position ({\em left}).}
\label{fig:azimuth}
\end{figure}

\end{appendix}
\end{document}